\documentclass[reprint,prl,nofootinbib,preprintnumbers]{revtex4-1}
 \usepackage{graphicx,psfrag,epsfig}
 \def\lsim{\mathrel{\vcenter{\hbox{$<$}\nointerlineskip\hbox{$\sim$}}}}
\def\gsim{\mathrel{\vcenter{\hbox{$>$}\nointerlineskip\hbox{$\sim$}}}}

\begin{document}
\preprint{PI-PARTPHYS-239, UM-DOE/ER/40762-506, UMD-PP-011-016}
\title{Sneutrino Dark Matter in Gauged Inverse Seesaw Models for Neutrinos}
%\vspace{0.5in}
\author {\bf  Haipeng An$^{1,2}$,  P. S. Bhupal Dev$^1$, Yi Cai$^3$, and R. N. Mohapatra$^{1}$\\}
%\vspace{0.3in}
%
\affiliation{\vspace{0.2cm}
$^{1}$Maryland Center for Fundamental Physics and
Department of Physics, University of Maryland, College Park, MD
20742, USA}
\affiliation{$^{2}$Perimeter Institute, Waterloo, Ontario N2L 2Y5, Canada}
\affiliation{$^{3}$Department of Physics, Shanghai Jiao Tong University, Shanghai 200240, China}
%\vspace{0.5in}
\begin{abstract}
Extending the Minimal Supersymmetric Standard Model (MSSM) to explain small neutrino masses via the inverse seesaw mechanism can lead to a new 
light supersymmetric scalar partner which can play the role of inelastic dark matter (iDM). It is a linear combination of the superpartners of the neutral fermions in the theory (the light left-handed neutrino and two heavy Standard Model singlet neutrinos) which can be very light with mass in $\sim 5-20$ GeV range, as suggested by 
some current direct detection experiments.  The iDM in this class of models 
has keV-scale mass splitting, which is intimately connected to the small 
Majorana masses of  neutrinos. We predict the differential scattering rate and 
annual modulation of the iDM signal which can be testable at future Germanium- 
and Xenon-based detectors. 
\end{abstract}

\maketitle

{\it Introduction -- }
A plethora of cosmological observations have clearly established the existence of a dark matter (DM) component to the matter budget of the Universe, which is about five times that of the visible matter  contribution. The mass and interaction properties of the DM are however not known, and experimental efforts are under way to detect it via its scattering against different kinds of nuclei, which will not only provide additional direct evidence for its existence, but will also reveal the detailed nature of its interaction with matter. Since  no particle in the Standard Model (SM) can play the role of DM, this will also be a sensitive probe of  physics beyond SM and supplement the new physics search at Large Hadron Collider (LHC). Supersymmetric extensions of SM provide one class of such models  with several natural candidates for DM (e.g. neutralino, gravitino, etc) when  $R$-parity symmetry is assumed.  However, 
if the DM mass turns out to be in the few GeV range, as suggested by some recent experiments~\cite{expts}, the minimal version (MSSM) may need to be extended 
to accommodate this, mainly due to its inability to reproduce the 
observed DM relic density~\cite{hooper}.

Another reason for considering extensions of MSSM is to explain small neutrino masses.  It would be interesting to see if the same extensions can also provide a new DM candidate and determine its properties. Simple ways to understand the smallness of neutrino masses are by adding one or more heavy SM singlet fermions to 
the MSSM . In these cases, the superpartner(s) of the singlet neutrino(s) with a small admixture of the left sneutrino can play the role of DM. 
%There exists a vast literature on such models with a single right-handed (RH) neutrino~\cite{one}, where one RH neutrino ($N$) per family is introduced for neutrino masses. 
In this paper, we focus on the supersymmetric inverse seesaw models~\cite{inverse} where one adds two SM singlet fermions $N$ and $S$.  In these models, there are three lepton number carrying electrically neutral fermions per family, namely $(\nu, N, S)$.  The DM particle is the lightest super-partner (LSP) of 
the model which can be a linear combination of the superpartners of $(\nu, N, S)$~\cite{valle,khalil,michaux,kang2011}. Current literature on the subject discusses two classes of such models. In the first class, inverse seesaw is considered within the framework of MSSM~\cite{valle}; however, in these models, one needs to omit terms in the superpotential that are allowed by the symmetries of the Lagrangian. The second class of inverse seesaw DM models extend the gauge symmetry to $SU(2)_L\times U(1)_Y\times U(1)_{B-L}$ so that the inverse seesaw mass matrix arises from a $B-L$ gauge symmetry~\cite{khalil}. However, the $B-L$ gauge symmetry discussed in Ref.~\cite{khalil} does not arise from a grand unified theory (GUT). Yet, a third class uses global $B-L$ symmetry to restrict the inverse seesaw matrix to the desired form~\cite{michaux}. 
%In this letter, we mainly discuss two examples, which are inverse seesaw extension in MSSM and in a SUSY left-right symmetric model (SUSYLR)~\cite{model}. 
In this letter, we extend the MSSM gauge group to the supersymmetric Left-Right (SUSYLR) 
gauge group $SU(2)_L\times SU(2)_R\times U(1)_{B-L}$ so that not only does the inverse seesaw matrix arise naturally 
but, as was shown in Ref.~\cite{model}, this model can emerge as a TeV scale theory from $SO(10)$ GUT implying different dynamical properties of the DM particle than the previous woks. 
%We also discuss the inverse seesaw extension in MSSM as a special case of this SUSYLR model and show that this kind of 
%SUSY inverse seesaw models with the sneutrino LSP can share common signatures in both DM direct detection experiments and LHC experiments. . 
%In addition it incorporates very different
%dynamics for the DM particle, as we see below.

Working within this new class of models, we find the following results: (i) A 
linear combination of the superpartners of the new SM singlet fermions  can be 
the LSP with mass from a few GeV up to about 100 GeV without running into 
conflict with known low energy observations. 
%Although getting this LSP in the 
%5-20 GeV range requires some fine tuning of parameters, we can still 
%get the right relic density. 
(ii) The $S$-fermion, which is given a small lepton number violating mass to understand small neutrino masses, leads to a splitting of the above complex scalar LSP into two closely-spaced real scalar fields, 
the lighter of which (we'll denote it by $\chi_1$) is the true DM field and is accompanied by its slightly heavier partner field ($\chi_2$) with a mass difference of order keVs. A consequence of this is that the direct detection process involves a dominantly inelastic scattering mode with the nucleus (${\cal N}$) where $\chi_1+{\cal N}({\cal A},{\cal Z})\to \chi_2+{\cal N}({\cal A},{\cal Z})$~\cite{neal}, and can therefore be tested in direct DM detection experiments~\cite{neal2}. The inelastic property arises naturally since the gauge Noether current coupling to the $Z$ (and $Z'$ in models with extended gauge symmetries) necessarily connects $\chi_1$ to $\chi_2$; also, any possible elastic contribution (mostly through the Higgs mediation) is highly suppressed due to small Yukawa couplings to light quarks. 
We believe this is an important point which has not been properly emphasized in earlier papers. 
(iii) The new TeV scale gauge dynamics in these models leads to new annihilation mechanisms for DM in the early Universe responsible for 
its current relic density. 
%(iv) Finally, we also note a collider signature which is specific to sneutrino combination rather than the neutralino being the DM.

{\it General Structure of Supersymmetric Inverse seesaw -- }
The inverse seesaw model~\cite{inverse} for neutrino masses uses the following mass matrix involving the $(\nu,N,S)$ fields:
\begin{equation}
{\cal M}_{\rm inv} = \left(\begin{array}{ccc}  0 ~&~ M_{D}^{T} ~&~ 0 \\
						   M_{D} ~&~ 0 ~&~ M_{R} \\
						   0 ~&~ M_{R}^{T} ~&~ \mu_{S}/2 \\ \end{array}\right)\ ,
						   \label{eq:1}
\end{equation}
where we have suppressed the family index. This leads to the neutrino mass formula $M_{\nu} = M_{D} M_{R}^{-1} \mu_{S} \left(M_{D}M_{R}^{-1}\right)^T$. Here the smallness of the 
neutrino masses arises from the small parameter $\mu_S$, which for $M_R\sim$ TeV and $m_D\sim $ GeV, has a value in the keV range.

Neglecting the keV-scale lepton number violation effect, the mass eigenstates are complex scalars in the basis of $(\widetilde\nu, \widetilde N^\dagger, \widetilde S)$, and the lightest sneutrino eigenstate can be written as
\begin{eqnarray}
	\widetilde \chi_{1} &=& \sum_{i=1}^{3}(U^{\dagger})_{1\nu_{i}} \widetilde \nu_{i} + (U^{\dagger})_{1 N_{i}} \widetilde N^{\dagger}_{i} + (U^{\dagger})_{1 S_{i}}\widetilde S_{i}
\label{eq:eigenstate}
\end{eqnarray}
where $U$ is a $9\times9$ unitary matrix that diagonalizes the full neutrino 
mass matrix given by Eq.~(\ref{eq:1}). We note here that since the entries in the $\widetilde{N}, \widetilde{S}$ sector of the sneutrino mass matrix are $\sim$ TeV, to get an LSP in the GeV-range, we do need some fine tuning; however, as shown 
in the next section, we are able to get the right relic abundance even with 
such low-mass DM. Also, from 
universality arguments, the model requires the sneutrino LSP to be always 
below $\sim 100$ GeV, beyond which the lightest neutralino becomes the LSP. 

When lepton number violation is 
invoked, the splitting terms $\sum_{m,n=1}^9 A_{mn}\widetilde\chi_m\widetilde\chi_n$ can be generated in the sneutrino sector (similar to those in Ref.~\cite{march}), and up to leading order in the lepton 
number violating mass term $\mu_S$, the mass splitting of LSP can be written as
\begin{equation}
\delta M_\chi = 4|A_{11}|/M_\chi\ ,
\end{equation}
where generically $A_{11}\sim \mu_S M_{\rm SUSY}$.  If $M_\chi$ is also of 
order of the SUSY breaking scale (assumed to be around TeV), the splitting is 
of order $\mu_S$, and if $M_\chi$ is much lower than $M_{\rm SUSY}$ as in some region of parameter space in SUSYLR, the splitting can be enhanced. 

The mass matrix in Eq.~(\ref{eq:1}) arises  
%in the MSSM from the following superpotential:
%\begin{equation}
%{\cal W}_{1} = {\cal W}_{\rm MSSM} +  Y_{\nu}NH_{u} L +  M_{R}N S + \frac{1}{2} S \mu_{S} S\ ,
%\end{equation}
in the  SUSYLR model~\cite{model}, after symmetry breaking, from the superpotential 
\begin{equation}
{\cal W} = {\cal W}_{\rm MSSM} +  Y_{\nu} L \Phi L^c+  Y_{S}S\phi_R L^c + \frac{1}{2} S \mu_{S} S\ ,
\end{equation}
where $\Phi$ is a bi-doublet under  $SU(2)_L\times SU(2)_R$ with $B-L$ charge zero, and $\phi_R$ is a right-handed doublet with a $B-L=1$
responsible for $B-L$ breaking. 

%In the MSSM version, since $S$ and $N$ are singlet fields, the gauge symmetry does not forbid terms like $L H_u S$ and $NN$. However, these two terms are also lepton number violating, so their sizes should be comparable with $\mu_S$. It is easy to see that the $NN$ term does not spoil the inverse seesaw structure since the rank of mass matrix is still two if only the $NN$ term is invoked. The $L H S$ term is a dimension four operator, and if lepton number is violated spontaneously, the Wilson coefficient of this operator can be written as $\mu_S/\Lambda$, where $\Lambda$ can be much larger than the electroweak symmetry breaking 
%scale $v_{\rm wk}=246$ GeV; therefore this term can also be suppressed and will not spoil the structure of inverse seesaw given by Eq.~(\ref{eq:1}). 
%These issues do not exist in the case of SUSYLR, 
%since the $SU(2)_R$ gauge symmetry automatically forbids such terms. 

\begin{figure}[h!]
	\includegraphics[width=2.49cm]{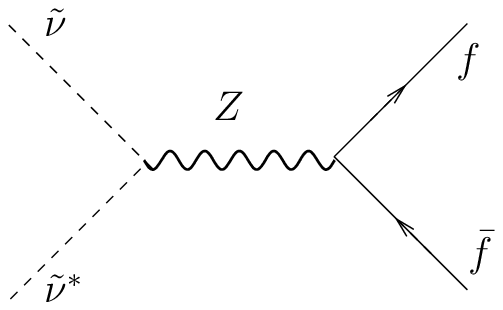}
	\includegraphics[width=2.49cm]{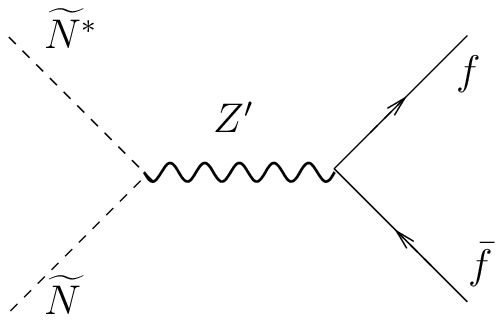}
	\includegraphics[width=2.49cm]{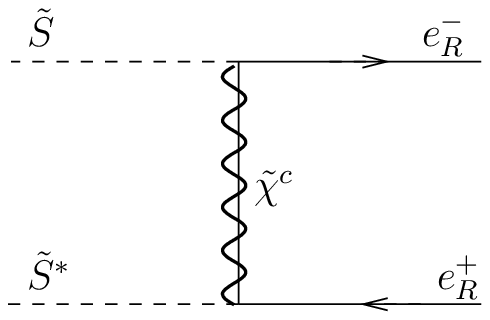}
	\caption{The dominant annihilation channels of the sneutrino DM in SUSYLR model.}
	\label{fig:annihilation}
\end{figure}

{\it Relic Density -- }
From Eq.~(\ref{eq:eigenstate}), we see that the sneutrino DM is a linear combination of the $(\widetilde{\nu},\widetilde{N}^\dag,\widetilde{S})$ fields. 
Therefore, as shown in Fig.~\ref{fig:annihilation}, its annihilation 
channels involve three major contributions, one from each component. 
Note that the second and third channels 
are the new contributions in SUSYLR. To leading order, the 
expressions for the annihilation cross 
sections in the $s$- and $t$-channel are respectively given by
\small\begin{eqnarray}
	\sigma_{s} &\simeq& \frac{g_{2L}^4\kappa_fN_c}
	{96\pi \cos^4\theta_W}\frac{sv}
	{(s-M_Z^2)^2+M_Z^2\Gamma_Z^2}\left[c_0^2+c_1^2\left(\frac{g_{2R}}{g_{2L}}\right)^4\right.
	\nonumber\\
	&&
\times \left.\left(\frac{\cos^{12}\theta_W}{\cos^2{2\theta_W}}\right)
\left(\frac{(s-M_Z^2)^2+M_Z^2\Gamma_Z^2}
	{(s-M_{Z'}^2)^2+M_{Z'}^2\Gamma_{Z'}^2}\right)\right.\nonumber\\
	&& \left. +
	2c_0c_1\left(\frac{g_{2R}}{g_{2L}}\right)^2\frac{\cos^8\theta_W}{\cos{2\theta_W}}\right.\nonumber\\
	&& \left.\times 
	\left(\frac{(s-M_Z^2)(s-M_{Z'}^2)+M_ZM_{Z'}\Gamma_Z\Gamma_{Z'}}
	{(s-M_{Z'}^2)^2+M_{Z'}^2\Gamma_{Z'}^2}\right)\right]\nonumber\\
	\sigma_t &\simeq& \frac{Y^4_S c_2^2}{96\pi}\frac{sv}
	{\left(M_{\widetilde\phi_R}^2-M_\chi^2\right)^2}
	\label{eq:stcross}
\end{eqnarray}
where $\kappa_{f} = (I_{3f}-Q_f\sin^2\theta_W)^2+(Q_f\sin^2\theta_W)^2$, 
$N_c = 3(1)$ for quarks (leptons), $v = \sqrt{1-4M_\chi^2/s}$ is the 
speed of the DM particle in the center-of-mass frame, and 
$c_{(0,1,2)} = \sum_{i=1}^3|U_{(\nu,N,S)_{i1}}|^2$. We note 
that both $s$- and $t$-channel annihilations in our case are 
$p$-wave scattering, as expected from symmetry arguments. 
For low-mass DM ($M_\chi <20$ GeV), and assuming $Y_S\sim {\cal O}(1)$ and 
$M_{\widetilde\phi_R}\lsim 500$ GeV, the $t$-channel 
involving only leptonic final states turns out to be the dominant 
contribution in our case. For this reason, we do not show the interference 
term between $s$- and $t$-channels for leptonic final states in 
Eq.~(\ref{eq:stcross}). 
 
\begin{figure}[h!]
\includegraphics[width=4.1cm]{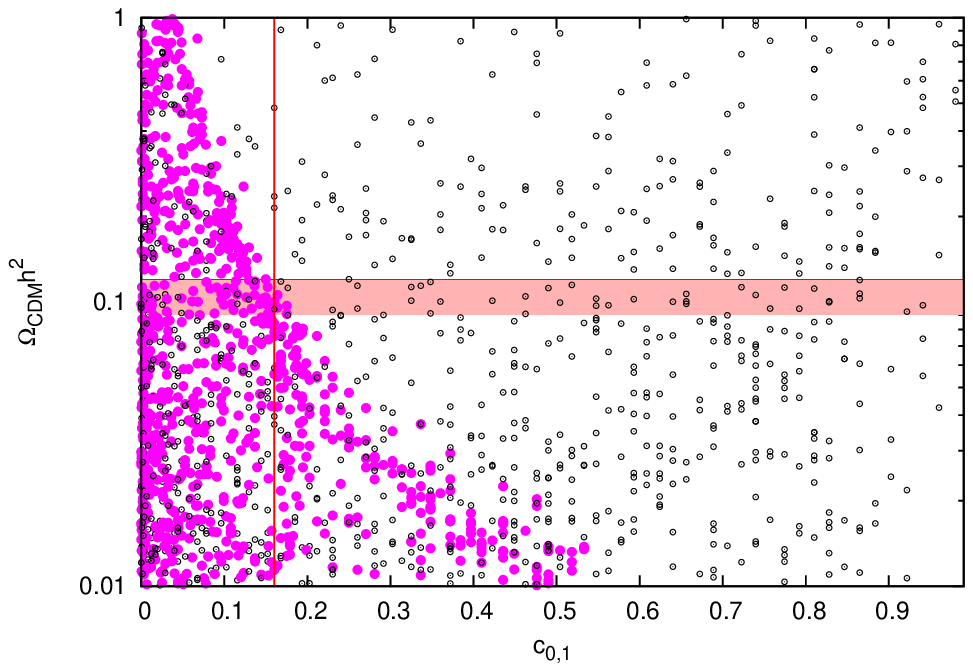}
\hspace{0.2cm}
\includegraphics[width=4.1cm]{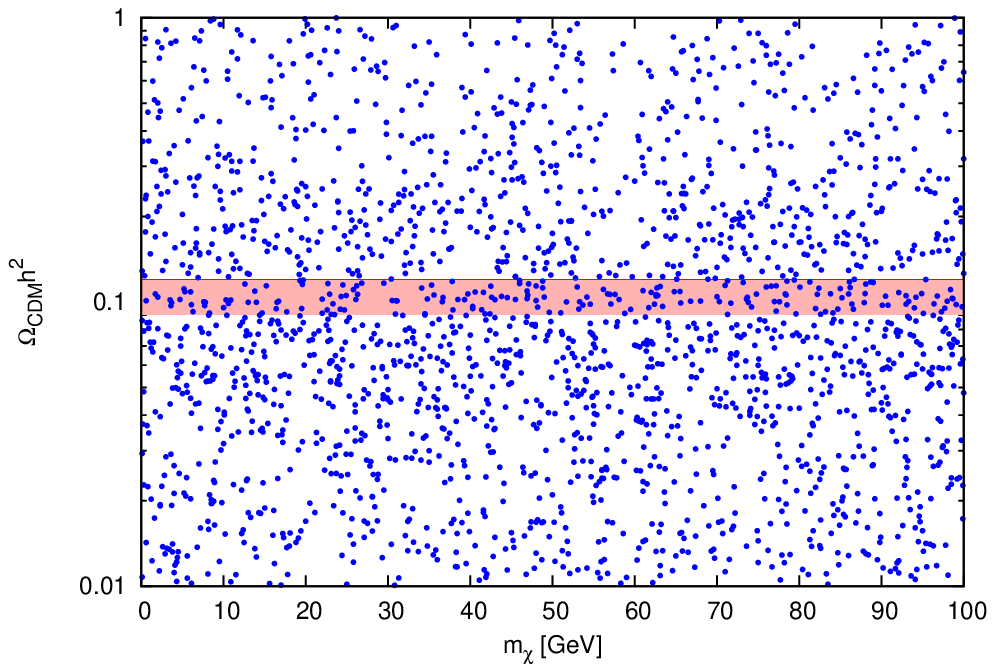}
\caption{In the left-panel, the purple (black) points correspond to 
the allowed values in the $c_{0(1)}$-relic density plane, for the mixed 
sneutrino DM in our model; the vertical line shows the upper limit for $c_0$ 
from invisible $Z$-width constraint. The 
right-panel shows the scatter plot of relic density prediction for light LSP 
mass. The horizontal shaded region is the $3\sigma$ limit obtained from 7-year 
WMAP data.} 
\label{fig:relic}
\end{figure}

The annihilation cross section for the $Z'$-channel is suppressed compared to 
the $Z$-channel by a factor $(c_1/c_0)^2(M_Z/M_{Z'})^4$. Also, we find that 
the correct DM relic density is obtained only for $c_0<0.16$, 
as shown in the left-panel of Fig.~\ref{fig:relic}. This also agrees with 
the invisible $Z$-decay width constraint (as shown by the vertical line in 
Fig.~\ref{fig:relic}). 
The right panel of Fig.\ref{fig:relic} shows the the scatter plot of the 
predictions for the LSP relic density; 
we find enough parameter range where the correct relic density is reproduced 
for a light DM. 

{\it Direct Detection -- }
As the sneutrino DM ($\widetilde{\chi}_1$) in inverse seesaw 
is a real scalar field accompanied by its slightly heavier partner field, 
it has both elastic and inelastic interaction with nuclei in direct detection 
experiments. 

The direct detection channel mediated by the SM-like Higgs boson is due to the 
interaction term $\lambda h_0 \widetilde{\chi}_1^\dag
\widetilde{\chi}_1$, where $\lambda$ is mainly 
the $D$-term contribution which can be simplified to 
%$(g_1^2 + g_2^2) v_{\rm wk} c_0/4$ for MSSM and  
$(g_{2L}^{2}c_0+g_{2R}^{2}c_1)v_{\rm wk}/4$ for SUSYLR 
assuming the decoupling limit of the MSSM Higgs mixing angles, i.e. large $\tan\beta$ and $\alpha\simeq \beta -\pi/2$~\cite{martin}. After invoking the lepton number violation effect, a mass splitting is generated between $\chi_1$ and $\chi_2$, and the interaction term can be rewritten as 
%\begin{equation}
$\frac{\lambda}{2} h_0(\chi_1^2 + \chi_2^2)$
%\ 
%\end{equation}
 which is clearly an elastic interaction. 

The direct detection channel conducted by gauge bosons can be written as 
\begin{eqnarray}
	&&\frac{i}{2\cos\theta_{W}}\left(c_0 g_{2L}Z^{\mu}+c_1\frac{\cos^2\theta_W}{\sqrt{\cos{2\theta_W}}}g_{2R}Z'^{\mu}\right)\nonumber\\
	& &~ ~ ~ ~ \times (\widetilde \chi_{1} \partial_{\mu} \widetilde \chi_{1}^{\dagger} - \widetilde \chi_{1}^{\dagger} \partial_{\mu} \widetilde \chi_{1}),
\end{eqnarray}
where $\theta_{W}$ is the Weinberg angle.
%, and in the MSSM version, only the first term exists (i.e. $c_1=0$). 
After the lepton number violation term is included, the 
interaction term is of the form 
$iZ^\mu(\chi_1 \partial_\mu \chi_2-\chi_2\partial_\mu\chi_1)$. 
Therefore, the collisions between $\chi_1$ and nucleus conducted by gauge 
bosons is inelastic. 

The differential scattering rate of DM particle on target nucleus in 
direct detection experiment can be written as 
\begin{equation}\label{dRdEr}
\frac{dR}{dE_{r}} = \frac{\rho_{\chi_1}}{M_{\chi}}\int_{|\textbf v|>v_{\rm min}}d^{3}\textbf v \frac{A^{2}_{\rm eff}\bar\sigma_{N}}{2\mu_{\chi N} |\textbf v|} F^{2}(|\textbf q|) f(\textbf v)\ ,
\end{equation}
where $E_{r}$ is the nuclear recoil energy, $\rho_{\chi_1}$ is the local mass density of DM, $M_\chi$ is the mass of the DM particle, $\sigma_{N}$ is the DM-nucleon cross section, and $\mu_{\chi N}$ is the reduced mass of DM and the target nucleus. $\bar\sigma$ and $A_{\rm eff}$ are defined as $\bar\sigma = (\sigma_{p}+\sigma_{n})/2$ and $A_{\rm eff} = \sum_{i\in {\rm isotopes}} 2 r_{i} [{\cal Z} \cos\theta_{N}+(A_{i}-{\cal Z})\sin\theta_{N}]^{2}$, where $r_{i}$ are relative abundances of isotopes, and $\tan\theta_{N} = {\cal M}_{n}/{\cal M}_{p}$, ${{\cal M}_{n,p}}$ being the DM scattering amplitudes off neutron and proton respectively~\cite{feng}. $F(|\textbf q|)$ is the nuclear form factor and $f(\textbf v)$ is the velocity distribution of the local galaxy. $v_{\rm min}$ is the minimal velocity needed to generate the nuclear recoil energy $E_{r}$, which can be written as~\cite{neal} 
\begin{equation}\label{vmin}
v_{\rm min} = \frac{1}{\sqrt{2M_{N} E_{r}}}\left(\frac{M_{N}E_{r}}{\mu_{\chi N}}+\delta\right)\,
\end{equation}
where $\delta$ is the mass gap, and $\delta = 0$ corresponds to the case of 
elastic scattering. 

In the case of elastic scattering conducted by Higgs, 
$\sigma_N$ in Eq.~(\ref{dRdEr}) is the total scattering cross section which 
can be written as 
\begin{eqnarray}
\sigma^{\rm el}_{N} &=& \frac{\lambda^{2}(M^{2}_{N}\sum_q\langle N|m_q \bar q q|N\rangle)^{2}}{4\pi v_{\rm wk}^{2}M_{h}^{4}(M_{N}+M_{\chi})^{2}}	\ ,
\label{eq:elastic}
\end{eqnarray}
whereas in the case of inelastic scattering, $\sigma_N$ can be written as
\begin{eqnarray}
	\sigma^{\rm in}_{p,n} = \frac{g_{2L}^{4}\kappa_{p,n}}
	{4\pi\cos^4\theta_W M_Z^4}
	 \frac{M_{p,n}^{2} M_{\chi_{1}}^{2}}{(M_{\chi_{1}}+M_{p,n})^{2}}
	 \left[c_0^2+c_1^2\left(\frac{g_{2R}}{g_{2L}}\right)^4
\right.
	 \nonumber\\
	 \left.\times 
	 \left(\frac{M_Z}{M_{Z'}}\right)^4\frac{\cos^{12}\theta_W}{\cos^2{2\theta_W}}+ 2c_0c_1\left(\frac{g_{2R}}{g_{2L}}\right)^2\left(\frac{M_Z}{M_{Z'}}\right)^2\frac{\cos^8\theta_W}{\cos{2\theta_W}}\right]\ ,\nonumber\\
	 \label{eq:inelastic}
\end{eqnarray}
where the first term in the bracket is induced by $Z$ boson whereas the second term is due to the $Z'$ boson in SUSYLR and the last term is due to the interference between the two. 
The factors $\kappa_p = \left(\frac{3}{4}-\sin^2\theta_W\right)^2$ and 
$\kappa_n=\left(\frac{3}{4}\right)^2$ are due to the different coupling of 
the vector bosons to proton and neutron respectively. 

\begin{figure}[h!]
	\includegraphics[width=8cm]{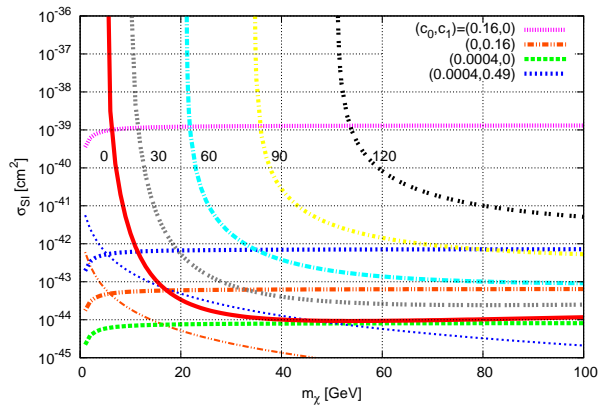}
	\caption{The model prediction for the scattering cross 
	section of the sneutrino DM off a nucleon as a function of 
	its mass, for various choices of the left and right sneutrino 
	components. Also shown are the 
	XENON100 constraints for mass gap $\delta= (0,~30,~60,~90,~120)$ keV  
	which were obtained by Feldman-Cousins method~\cite{feldman} 
	using the 100 days of live data~\cite{xenon}.}
	\label{fig:mDM_sigma}
\end{figure}

It is important to note here that for large fractions of the left 
sneutrino component ($c_0\gsim 10^{-3}$), the scattering is mostly dominated by the $Z$-exchange, and 
 is hence inelastic. The $Z'$ contribution to the inelastic channel is always 
suppressed by its mass, and similarly, the elastic cross section is 4-5 orders 
of magnitude smaller than the $Z$-dominated inelastic contribution because the 
coupling of the Higgs to nucleon is suppressed by light quark masses. However, 
for $c_0\lsim 10^{-4}$ and large $c_1$, the inelastic contribution, 
suppressed by the $Z'$-mass, could become comparable to the 
elastic counterpart. This is shown in
Fig.~\ref{fig:mDM_sigma}, where we have plotted the cross section as a function 
of the DM mass for various values of $c_0$ and $c_1$. One can see that for  
small $c_0$ and large $c_1$ values (blue and orange curves), the 
cross section is dominated by the elastic channel (thin lines) for low mass DM 
and by inelastic channel mediated by $Z'$ for $M_\chi\gsim 10$ GeV (thick lines), whereas for 
small or zero $c_1$ component (pink and green curves), the scattering is always dominated by the inelastic channel mediated by $Z$. It is interesting to 
note here that the current XENON100 data constrains most of the parameter space 
of the model and for large $c_0$, puts an upper bound on the DM mass. 
This can be seen clearly from Fig.~\ref{fig:mDM_sigma} where we have shown 
the XENON100 limits on scattering cross section for various values of the 
mass gap, starting from zero on the left (red solid curve, corresponding to the 
elastic case) to $\delta=30,~60,~90$ and 120 keV cases. 
We note that for small mass gaps of order a few keV (as expected in this 
model), the XENON100 constraints leave only the low mass iDM 
(below 20 GeV) open in this model. Similar 
constraints on the iDM mass can be obtained using the existing data from other 
direct detection experiments~\cite{otherex}, but the XENON100 constraints are 
found to be the most stringent. 

The nuclear recoil energy in both elastic and inelastic scattering has a 
maximum determined by the escape velocity of DM in the local galaxy, and  
a minimum determined by the mass gap for inelastic case. Therefore, the topology of the differential scattering rate for inelastic scattering is very different from the elastic scattering, which can be used to determine whether the DM is 
inelastic or not. Furthermore, for inelastic scattering, if the mass gap is comparable to the kinetic energy of DM, for certain nuclear recoil energy, due to Eq.~(\ref{vmin}), the required velocity is pushed to the tail end of the 
velocity distribution where the motion of the earth has a larger effect and therefore the DM annual modulation signal gets enhanced. 

\begin{figure}[h!]
	\includegraphics[width=4.0cm]{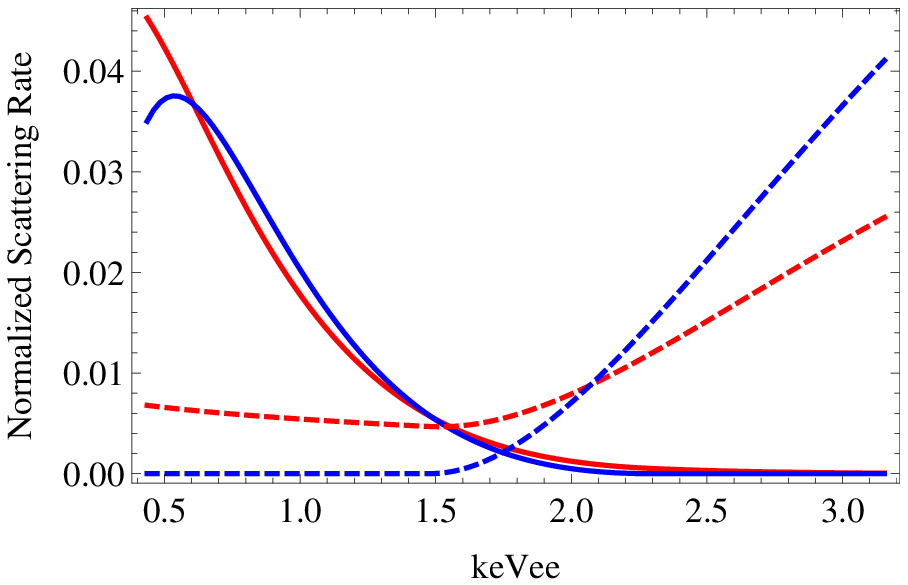}
	\hspace{0.2cm}
	\includegraphics[width=4.0cm]{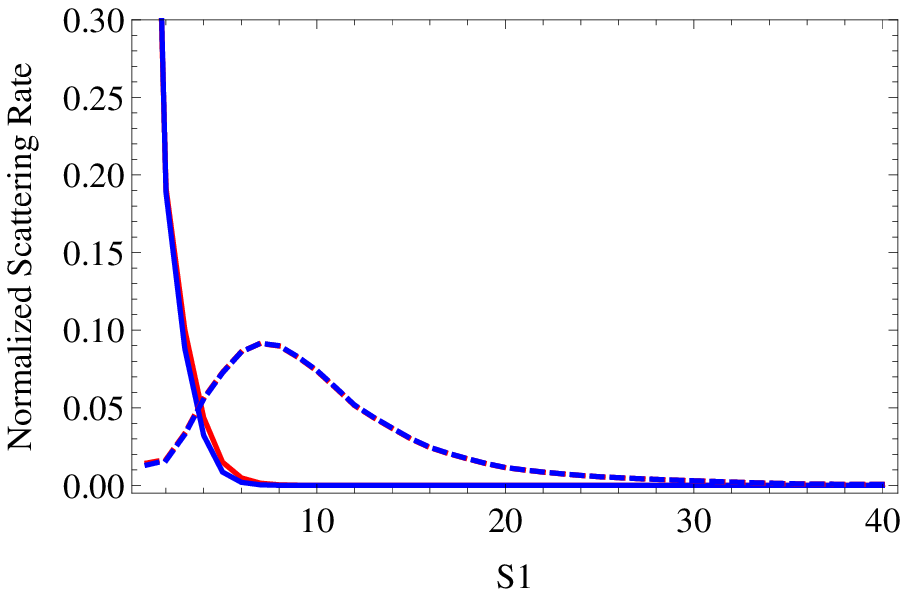}\\
	\vspace{0.2cm}
	\includegraphics[width=4.0cm]{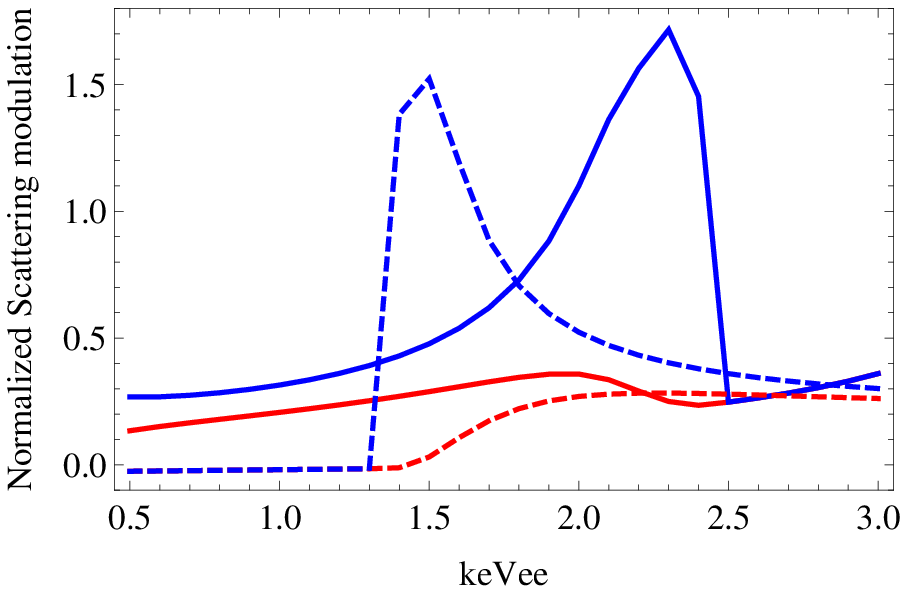}
	\hspace{0.2cm}
	\includegraphics[width=4.0cm]{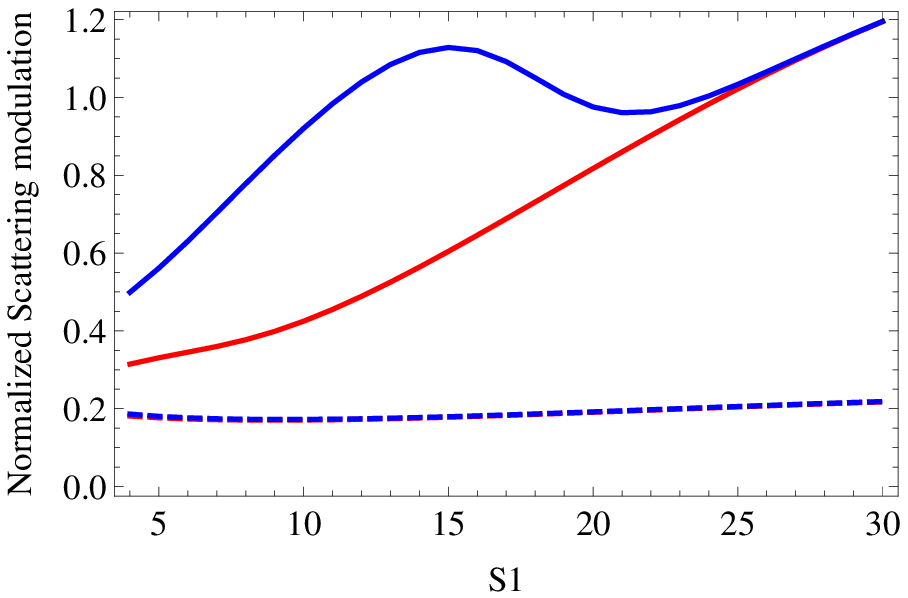}
	\caption{The model prediction for differential scattering rates 
	(normalized to one) and annual modulations for germanium and xenon detectors. The upper two plots show the 
	normalized scattering rate for germanium (left) and xenon (right) 
	detectors. The lower two plots show the annual modulation in the same 
	detectors. Red and blue curves are for $(c_0,c_1) = (0.001,0.1)$ and $(c_0, c_1)=(0.1,0.001)$, respectively, while the solid and dashed curves are for $(M_{\chi},\delta) = (10 {\rm GeV}, 20 {\rm keV})$ and  $(50 {\rm GeV}, 60 {\rm keV})$, respectively.}
	\label{fig:prediction}
\end{figure}

The predicted normalized differential scattering rate and annual modulation for Germanium- and Xenon-based detectors are shown in Fig.~\ref{fig:prediction} with different choices for parameters of the SUSYLR model. The mass of $Z'$ is taken to be 1 TeV in both cases. The red and blue curves are for $(c_0,c_1) = (0.001,0.1)$ and $(c_0, c_1)=(0.1,0.001)$, corresponding to the $Z'$ and $Z$ dominance, 
respectively, in the inelastic scattering between DM and target nucleus. 
The latter case is similar to the MSSM version of inverse seesaw. To translate 
the differential rates to experimental quantities, namely the 
electron equivalent 
recoil energies in germanium detector and the $\rm S_1$ signal in xenon 
detector, we have used the quenching factor and scintillation efficiencies from CoGeNT~\cite{cogent} and XENON100~\cite{xenon} experiments respectively.
 One can see from the first two plots in Fig.~\ref{fig:prediction} that a peak 
 shows up if the scattering is dominated by inelastic interactions. 
 Furthermore, one can see from the third and fourth plots that 
 for inelastic case, the annual modulation can be larger than 100\% in some 
 energy region. Also from the solid blue curves in the first and third plots,  
 one can see that the energy regions for large recoil energy and large 
 modulation can be separated from each other, which provides a chance to 
 fit the anomaly observed by the CoGeNT experiment. In these plots, 
 $A_0$ and $A_1$ are the zeroth and first Fourier modes of the differential 
 scattering rate; for a detailed definition, see Ref.~\cite{An:2011ck}. 

%, especially 
%with XENON100 constraints, in an iDM model, 
%even though there still exists a narrow parameter space to do so~
%\cite{impurity}. 
We note here that it is also possible to explain the recent CRESST-II data for 
an iDM with mass splitting ${\cal O}(100)$ keV, consistent with the null 
results from XENON100~\cite{zupan}. However, it is difficult to do so for 
DAMA/LIBRA, even though there still exists a narrow 
iDM parameter space that does fit DAMA/LIBRA data with CRESST-II~
\cite{impurity}. For this reason, we did not include the DAMA/LIBRA results 
in our discussion on annual modulation. 

{\it Comments --} 
As noted earlier, there exists a domain in the parameter space for which 
the dominant annihilation channel of the DM is to 
lepton-anti-lepton final states. The relative branching fractions depend 
on the masses of the RH neutrinos and are somewhat model-dependent. 
If we take the model in Ref.~\cite{model} 
as a guide, we expect them to be of similar order. These dominantly-leptonic 
annihilation modes can, in principle, be important in 
understanding signals from the galactic center, radio filaments as well as WMAP haze~
\cite{hooper2}; however, we expect these effects to be suppressed in our model 
due to the $p$-wave nature of the dominant DM annihilation channels. 

Furthermore, we observe that there exist upper limits on the direct DM 
detection 
cross section from monojet plus missing energy 
searches in colliders~\cite{monojet}. However, 
for light spin zero dark matter only weakly interacting with the nucleons, 
the collider limits on the spin-independent cross 
sections are weaker than the direct search bounds. 

Also, depending on the sparticle spectrum prediction of the model, the 
sneutrino LSP can have distinctive collider signatures. In particular, if  
the gluino is lighter (heavier) than the squarks, the 
sneutrino LSP can have interesting four (two) jet+like (opposite)-sign 
dilepton signals with missing energy~\cite{lessa}. A detailed collider 
simulation of these events for our model will be given elsewhere~\cite{ABCM}.

{\it Conclusion -- }
To summarize, we have shown that the supersymmetric inverse seesaw model for 
neutrinos naturally leads to an 
inelastic scalar DM. The differential scattering rate and annual 
modulation predicted in these models might be tested in future direct detection experiments. 
The DM particle mass is found to be 
strongly constrained by the current XENON bounds, and for keV-scale mass 
splitting for the real scalar LSP states (as required by neutrino mass 
constraints), we find an upper limit of around 20 GeV on the DM mass. 
This is consistent with the model prediction for the sneutrino LSP mass which 
is required to be below $\sim 100$ GeV from universality arguments.  

Therefore, we might be able to identify SUSY inverse seesaw if from the 
collider search, we can confirm that the sneutrino is a long-lived particle, and from direct detection experiments, we observe an inelastic WIMP from the 
differential scattering rate and the annual modulation.

{\it Acknowledgment -- }  H. A. and Y. C. would like to thank X.~G. He 
for helpful discussions. H. A. is supported by the U. S. Department of Energy 
via grant DE-FG02-93ER-40762 and the research at Perimeter Institute is 
supported in part by the Government of Canada through NSERC and by the 
Province of Ontario through MEDT. The work of B. D. and R. N. M. is supported 
by National Science Foundation grant number PHY-0968854. Y. C. is supported 
by NSC, NCTS, NNSF and SJTU Innovation Fund for Post-graduates and Postdocs.

\end{document}